# Classification of Heterogeneous Operating Systems

**Kamlesh Sharma\* and Dr. T. V. Prasad\*\***
\* Research Scholar, \*\* Dean of Academic Affairs
Dept. of Comp. Sc. & Engg., Lingaya's University, Faridabad, India
Email: kamlesh0581@gmail.com, tvprasad2002@yahoo.com

*Abstract* — **Operating system is a bridge between system and user. An operating system (OS) is a software program that manages the hardware and software resources of a computer. The OS performs basic tasks, such as controlling and allocating memory, prioritizing the processing of instructions, controlling input and output devices, facilitating networking, and managing files. It is difficult to present a complete as well as deep account of operating systems developed till date. So, this paper tries to overview only a subset of the available operating systems and its different categories. Operating systems are being developed by a large number of academic and commercial organizations for the last several decades. This paper, therefore, concentrates on the different categories of operating systems with special emphasis to those that had deep impact on the evolution process. The aim of this paper is to provide a brief timely commentary on the different categories important operating systems available today.**

## I. INTRODUCTION

An operating system is a software that manages all the resources of a computer, both hardware and software, and provides an environment in which a user can execute programs in a convenient and efficient manner [1]. However, the principles and concepts used in the operating systems were not standardized in a day. In fact, operating systems have been evolving through the years [2]. There were no operating systems in the early computers. In those systems, every program required full hardware specification to execute correctly and perform each trivial task, and its own drivers for peripheral devices like card readers and line printers. The growing complexity of the computer hardware and the application programs eventually made operating systems a necessity. Initially, operating systems were not fully automatic as Hansen [3] defined an operating system as a set of manual and automatic procedures that enable a group of people to share a computer installation efficiently. It is fortunate enough for today's computer users that modern operating systems are fully automatic. Over the years, sustained research in operating systems gave rise to many novel concepts and ideas. Operating systems exist even today because they offer a reasonable way to solve the problem of creating a usable computing system [1]. Moreover, sophisticated operating systems increase the efficiency and consequently decrease the cost of using a computer [5]. A large number of operating systems of various types are available for both research and commercial purposes, and these operating systems vary greatly in their structures and functionalities [1].

Computers have progressed and developed so have the operating systems. Below is a basic list of the different operating systems and a few examples of operating systems that fall into each of the categories. Many computer operating systems will fall into more than one of the below categories.With the integration of computers and telecommunications, the mode of information access becomes an important issue. The designs of the prevalent human machine interfaces are more suitable for easier interpretation of information by computers than by human beings. The concept of machine being able to interact with people in a mode that is natural as well as convenient for human beings is very appealing.

## II. TYPES OF OPERATING SYSTEMS

Operating systems can be differentiated based on different parameters used by the existing operating systems of computers.
- Interface: CUI, GUI, TUI, VUI, HSUI
- Internal Architecture: By kernel type
- Mode: Batch Processing Operating System, Real Time Operating System, Single User, Single Tasking Operating System, Single User, Multi-Tasking Operating System, Multi-User Operating System, Distributed Operating System

## III. CLASSIFICATION BASED ON INTERFACE

The manner in which users interact with a program is known as its user interface. The user interface controls how data is entered and how information is displayed. There are mainly five types of user interfaces:

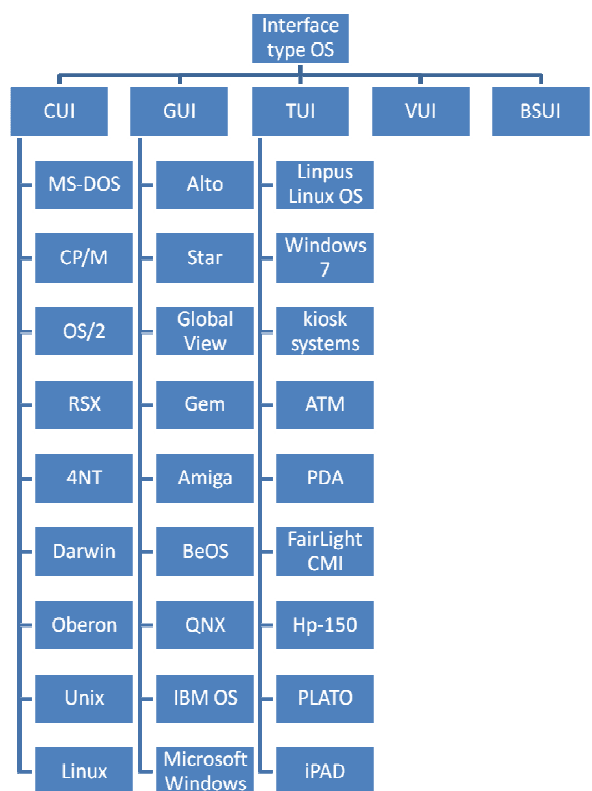

**Figure 1. Different Categories of OS with Examples**

*A. CUI (Command user Interface) or CLI (Command Line interface)*

The command line interface requires the use of the keyboard to enter commands to the computer. All commands are enter at the prompt and require exact spelling otherwise an error message will be displayed. This method of instructing a computer to perform a given task is referred to as "entering" a command: the system waits for the user to conclude the submitting of the text command by pressing the "Enter" key (a descendant of the "carriage return" key of a typewriter keyboard). A command-line interpreter then receives, parses, and executes the requested user command.

**Table 1. List of Command Line user interface**

| S. No | Name | Developed by | Year |
|---|---|---|---|
| 1 | MS-Dos | Microsoft Corporation | 1981 |
| 2 | IBM PC Dos | IBM and Microsoft | 1980 |
| 3 | CP/M CCP | Digital Research, Inc. / Gary Kildall | 1983 |
| 4 | DR-DOS | Digital Research | 1976 |
| 5 | Novell DOS | Digital Research | 1976 |
| 6 | OS/2 | IBM and Microsoft | 1987 |
| 7 | DEC`s RSX | Digital Equipment Corporation | 1972 |
| 8 | RSTS | Digital Equipment Corporation | 1970 |
| 9 | 4DOS for DOS | JP Software | 2004 |
| 10 | 4OS2 for OS/2 | JP Software | 1988 |
| 11 | 4NT or Take Command for windows | JP Software | 1988 |
| 12 | Windows PowerShell | Microsoft Corporation | 2006 |
| 13 | Darwin | Apple Inc. | 2000 |
| 14 | Oberon | Niklaus Wirth, Jürg Gutknecht | 1985 |
| 15 | Unix shells ( bash) | Brian Fox | 1989 |
| 16 | cmd.exe in Windows 7 | Microsoft corporation | 2009 |
| 17 | Mac OS X Terminal | Apple Inc. | 2008 |
| 18 | Linux | Many | 1991 |
| 19 | Unix Shell | Ken Thompson, Dennis Ritchie, Brian Kernighan, Douglas McIlroy, Joe Ossanna at Bell Labs | 1969 |
| 20 | RT-11 running on UKNC | Digital Equipment Corporation and Mentec Inc. | 1970 |

*B. GUI (Graphical User Interface)*

Graphical User Interface allow users to enter commands by pointing and clicking on icons, buttons, menu items and other objects with a mouse or other pointing devices. Programs running within a graphical environment are executed in a rectangular box called a window. GUIs can be used in computers, hand-held devices such as MP3 players, portable media players or gaming devices, household appliances and office equipment. A *GUI* represents the information and actions available to a user through graphical icons and visual indicators such as secondary notation, as opposed to text-based interfaces, typed command labels or text navigation. The actions are usually performed through direct manipulation of the graphical elements. Unlike a command line operating system like UNIX or MS-DOS, GUI Operating Systems are much easier for end-users to learn and use because commands do not need to be known or memorized. Because of their ease of use, GUI Operating Systems have become the dominant operating system used by end-users today.

**Table 2. List of Graphical user interface**

| S. No | Name | Developed by | Year |
|---|---|---|---|
| 1 | Xerox Alto | Xerox | 1973 |
| 2 | Xerox Star | Xerox | 1977 |
| 3 | Xerox Global View 2.1 | Xerox | 1996 |
| 4 | Xerox GlobalView for X | Xerox | 1992 |
| 5 | Xerox Rooms for X Windows | Xerox | 1992 |
| 6 | Three Rivers / ICL Perq | Three River Computer | 1979 |
| 7 | VisiCorp Visi On | IBM Inc. | 1982 |
| 8 | GEM 1.1 | Digital Research | 1984 |
| 9 | GEM 2.0 | Apple Computer | 1985 |
| 10 | GEM 3.11 |  | 1988 |
| 11 | Atari TOS 1.0 | Atari ST and TT series | 1985 |

| S. No | Name | Developed by | Year |
|---|---|---|---|
| | | of Computer | |
| 12 | Tandy Deskmate 3.69 | Tandy Radio | 1984 |
| 13 | GEOS For the Commodore 64 | Berkley Softworks | 1985 |
| 14 | DESQview/X | Quarterdeck | 1985 |
| 15 | AmigaOS 3.5 | Amiga International Inc. | 1985 |
| 16 | RISC OS 3 | RISCOS Ltd. | 1992 |
| 17 | RISC OS 4 | RISCOS Ltd. | 1999 |
| 18 | BeOS | BE Inc. | 1991 |
| 19 | QNX Demo Disk | QNX Software Systems | 1982 |
| 20 | Microsoft OS/2 V1.3 | Microsoft and IBM | 1987 |
| 21 | IBM OS/2 2.0 | IBM | 1992 |
| 22 | OS/2 Warp 3 | Microsoft | 1994 |
| 23 | IBM OS/2 Warp 4 | Microsoft and IBM | 1996 |
| 24 | eComStation Demo CD | Seresnity Systems | 2001 |
| 25 | Apple Lisa | Apple Computer Inc. | 1986 |
| 26 | Apple Desktop II | Apple Computer | 1996 |
| 27 | GS/OS for the Apple IIgs | Apple Computer | 1983 |
| 28 | Quark Catalyst 3.0 | Quark Incorporation | 1982 |
| 29 | Apple Macintosh | Apple Inc. | 1984 |
| 30 | Mac OS 7.5.5/7.6 | Apple computer | 1997 |
| 31 | Mac OS 8.1 | Apple computer | 1997 |
| 32 | Mac OS 9.2.2 | Apple Computer Inc. | 1999 |
| 33 | At Ease | Apple Computer | 1998 |
| 34 | At Ease for Workgroups | Apple Computer | 1988 |
| 35 | OPENSTEP 4.2 | Sun Microsoft | 1993 |
| 36 | Rhapsody Developer Release 2 | Apple computer | 1998 |
| 37 | MacOS X 10.1 | Apple inc. | 2001 |
| 39 | Mac OS X 10.4.6 (Tiger) | Apple Inc. | 2005 |
| 40 | Mac OS X 10.5 (Leopard) | Apple Inc. | 2007 |
| 41 | Linspire Five-O | Linspire Inc. | 2007 |
| 42 | Mandrake Linux 9.0 | | |
| 43 | Red Hat Linux 8.0 With GNOME/ Nautilus 2.06 | Red Hat | 2004 |
| 44 | IRIX 6.5 | Silicon Graphics, Inc. | 1998 |
| 45 | Wine | | |
| 46 | Fedora 7 GNOME | | |
| 47 | Ubuntu Linux 11.10 | Canonical Ltd. | |
| 48 | gOS 2.0.0-beta1 | Good OS LLC. | |
| 49 | NetBSD 5.0 | Berkeley Software Distribution | |
| 50 | SunOS | Sun Microsystem | 1999 |
| 51 | Suntools / SunView - SunOS 3.5 | | |
| 51 | Solaris | Sun Microsystem | 1993 |
| 52 | GlobalView | Xerox | 1992 |
| 53 | X Windows System | | |
| 54 | ReactOS | | |
| 55 | 98Lite | Shane Brooks | 1998 |
| 56 | Windows 1.0 | Microsoft | 1985 |
| 57 | Windows 2.0 | Microsoft | 1987 |
| 58 | Windows 3.0 | Microsoft Windows | 1990 |
| 59 | Windows 3.1x | Microsoft | 1992 |
| 60 | Windows 95 | Microsoft | 1995 |
| 61 | Windows 98 | Microsoft | 1998 |
| 62 | Windows ME | Microsoft | 2000 |
| 63 | Windows NT | Microsoft | 1993 |
| 64 | Windows 2000 | Microsoft | 2000 |
| 65 | Windows XP | | |
| 66 | Windows Server 2003 | Microsoft | 2003 |

*C. TUI (Touch screen User Interface)*

A touchscreen is an electronic visual display that can detect the presence and location of a touch within the display area. The term generally refers to touching the display of the device with a finger or hand. Touchscreens can also sense other passive objects, such as a stylus. Touchscreens are common in devices such as all-in-one computers, tablet computers, and smartphones.

**Table 3. List of Touch Screen User Interface**

| S. No | Name | Developed by | Year |
|---|---|---|---|
| 1 | Linpus Linux OS | | |
| 2 | Windows 7 | Microsoft | 2009 |
| 3 | Capacitive touch screen | E.A. Johnson | 1972 |
| 4 | Kiosk systems | University of Illinois at Urbana-Champaign | 1977 |
| 5 | Point of sale systems | McDonald's | 1974 |
| 6 | ATM | John Shepherd-Barron | 1967 |
| 7 | PDA | Psion | 1986 |
| 8 | Fairlight CMI | Peter Vogel and Kim Ryrie | 1979 |
| 9 | HP-150 | Hewlett-Packard | 1983 |
| 10 | PLATO IV | University of illinois | 1970 |
| 11 | iPAD | Apple Inc | 2010 |
| 12 | Smartphone | IBM, Nokia, Microsoft | 1992-2000 |

*D. VUI (Voice User Interface)*

A Voice–user interface (VUI) makes human interaction with computers possible through a voice/speech platform in order to initiate an automated service or process. A VUI is the interface to any speech application. Controlling a machine by simply talking to system. VUIs have become more commonplace, and people are taking advantage of the value that these hands-free, eyes-free interfaces provide in many situations.

E. BSUI (Brain Signal User Interface)

## IV. CLASSIFICATION BASED ON INTERNAL ARCHITECTURE

The internal architecture of Operating Systems are.
- Monolithic Kernel
- Microkernel Kernel
- Hybrid System
- Nanokernel
- Exokernel

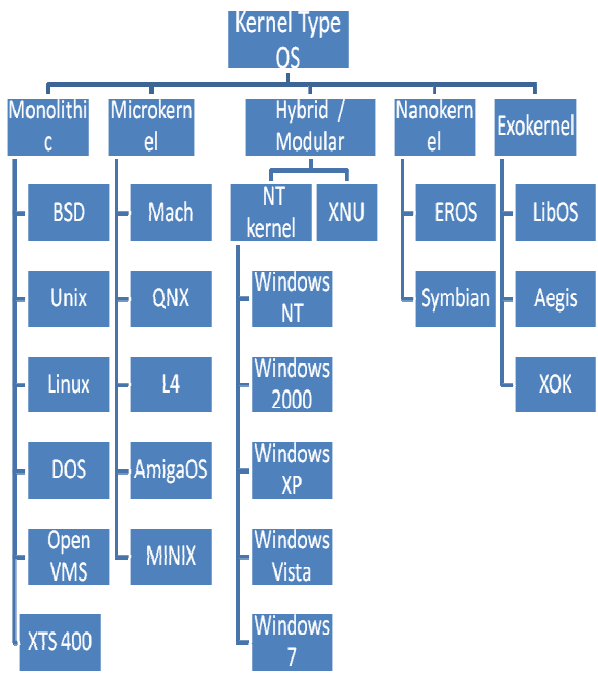

**Figure 2. Different Categories of OS based on internal Architecture with Examples**

*A. Monolithic Kernel*

A monolithic kernel is an operating system architecture where the entire operating system is working in the kernel space and alone as supervisor mode. The OS is written as a collection of Procedures, each of which can call any of the objects whenever it is needed. Each Procedure in the system has a well define interface in terms of parameters and results, is free to call any other one. The instruction switch machine from user mode to kernel mode and transfer control to the operating system. [13] [14]

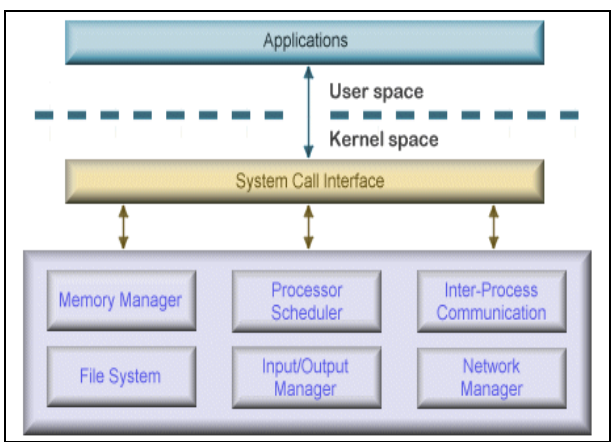

**Figure 3. Monolithic Kernel Architecture [22]**

*B. Microkernel Architecture*

Microkernel architecture includes only a very small number of services within the kernel in an attempt to keep it small and scalable. The services typically include low-level memory management, inter-process communication and basic process synchronization to enable processes to cooperate. [14] Its designs, most operating system components, such as process management and device management, execute outside the kernel with a lower level of system access. Kernels larger than 20,000 lines are generally not considered microkernel. [16][17]

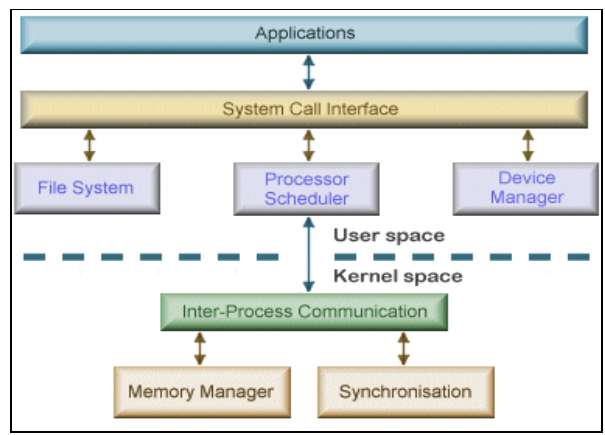

**Figure 4. Microkernel Architecture [22]**

*C. Hybrid Kernel(Macrokernel) Architecture*

A hybrid kernel is a kernel architecture based on combining aspects of microkernel and monolithic kernel architectures used in computer operating systems. The idea behind this category is to have a kernel structure similar to a microkernel, but implemented in terms of a monolithic kernel. In contrast to a microkernel, all operating system services are in kernel space. While there is no performance overhead for message passing and context switching between kernel and user mode, as in monolithic kernels, there are no performance benefits of having services in user space, as in microkernel.

*D. Nanokernel Architecture*

A kernel is a very small kernel where the total amount of kernel code, executing in the privileged mode of the hardware. [3] The term picokernel was sometimes used to further emphasize small size. It was a sardonic response to Mach, which claimed to be a microkernel while being monolithic, essentially unstructured, and slower than the systems it sought to replace. Subsequent reuse of and response to the term, including the picokernel coinage, suggest that the point was largely missed. Both Nanokernel and picokernel have subsequently come to have the same meaning expressed by the term microkernel. A virtualization layer underneath an operating system, this is more correctly referred to as a hypervisor. A hardware abstraction layer that forms the lowest-level part of a kernel, sometimes used to provide real-time functionality to normal OS's, likes Adeos. [6]

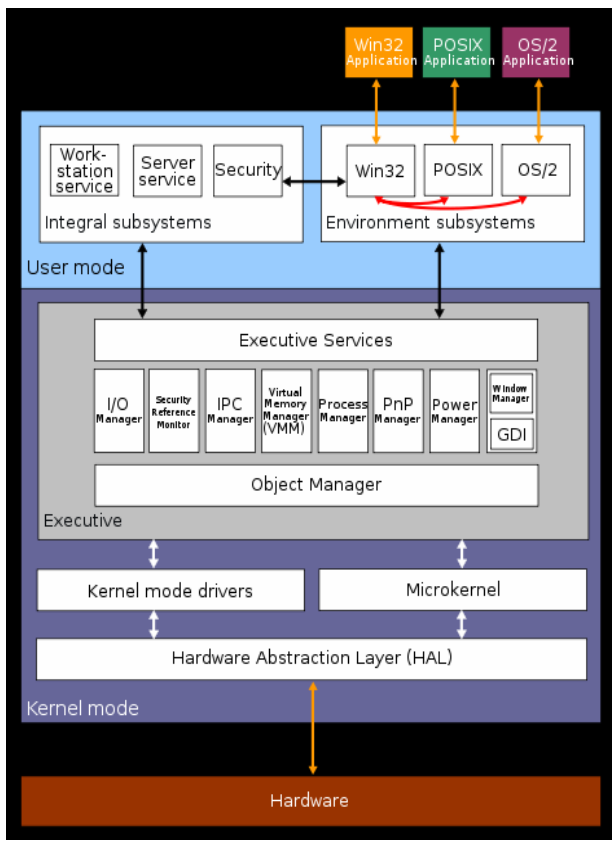

**Figure 5. Hybrid Kernel Architecture [23]**

*E. Exokernel Architecture*

Exokernel is tiny, since functionality is limited to ensuring protection and multiplexing of resources, which are vastly simpler than conventional microkernel's' implementation of message passing and monolithic kernels' implementation of abstractions. The idea behind Exokernel is to force as few abstractions as possible on developers, enabling them to make as many decisions as possible about hardware abstractions. Exokernel can be seen as an application of the end-to-end principle to operating systems, in that they do not force an application program to layer its abstractions on top of other abstractions that were designed with different requirements in mind. For example, in the MIT Exokernel project, the Cheetah web server stores preformatted Internet Protocol packets on the disk, the kernel provides safe access to the disk by preventing unauthorized reading and writing, but how the disk is abstracted is up to the application or the libraries the application uses.

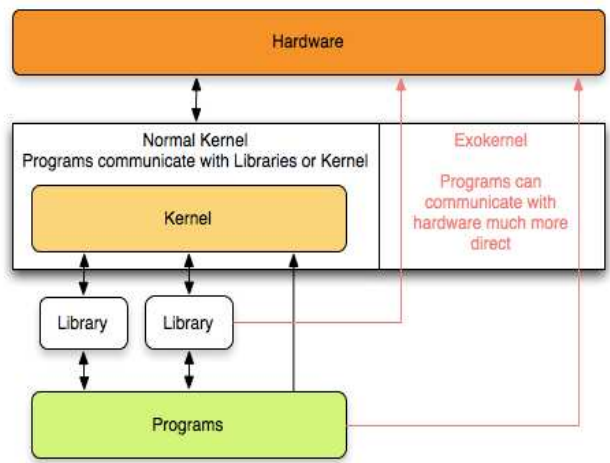

**Figure 6. Exokernel Architecture [24]**

V. CLASSIFICATION BASED ON PROCESSING CAPABILITY

Known modes of Operating Systems are
- Batch Processing Operating System
- Real Time Operating System
- Single User, Single Tasking Operating System
- Single User, Multi-Tasking Operating System
- Multi-User Operating System
- Distributed Operating System

*A. Batch Processing Operating System*

Interactions between the user and processor are limited in batch processing OS or there is no interaction at all during the execution of work. Data and programs that need to be processed are bundled and collected as a 'batch' and executed together. Batch processing OS are ideal in situations where the large amounts of data to be processed or similar data needs to be processed or similar processing is involved when executing the data. The system is capable of identifying times when the processor is idle at which time 'batches' maybe processed. Processing is all performed automatically without any user intervention. e.g.: SCOPE, KRONOS, NOS and EXEC [18]

*B. Real-time Operating System*

Real-Time OS which responds to inputs, immediately and generates results, instantly. This type of system is usually used to control scientific devices or complex systems that require a lot of processing like machinery and industrial systems and similar small instruments where memory and resources are crucial and constricted. [5] This type of devices have very limited or zero-end user utilities, so more effort goes into making the OS really memory efficient and fast (less coding), so as to minimize the execution time, in turn saving on power as well. e.g.:

VxWorks, PikeOS, eCos, QNX, MontaVista Linux and RTLinux. Windows CE, 8086 etc.

### C. Single User, Single tasking Operating System

Single-user OS are usable by a single user at a time. Being able to have multiple accounts on a Windows operating system does not make it a multi-user system. This type of OS is better version of Real time OS, where one user can do effectively one thing at a time, which means that doing more than one thing at a time is difficult in this type of OS. For instance: The palm OS in palm hand held computer is an example of single-task OS.

### D. Single user, Multi-Tasking Operating System

It allows more than one program to run concurrently like printing, scanning, word processing etc. e.g. MS Windows and Apple's Mac OS. Several applications maybe simultaneously loaded and used in the memory, while the processor handles only one application at a particular time it is capable of switching between the applications effectively to apparently simultaneously execute each application. Lots of operating system is seen everywhere today and is the most common type of OS, the Windows operating system would be an example. [19]

### E. Multi-User Operating System

It allows multiple users to simultaneously use the system, the processor splits its resources and handles one user at a time, the speed and efficiency at which it does this makes it apparent that users are simultaneously using the system, some network systems utilize this kind of operating system. Unix, VMS and mainframe operating systems, such as MVS, are examples of multi-user operating system.

### F. Distributed Operating System

In a distributed system, software and data maybe distributed around the system, programs and files maybe stored on different storage devices which are located in different geographical locations and maybe accessed from different computer terminals. While we are mostly accustomed to seeing multi-tasking and multi-user operating systems, the other operating systems are usually used in companies and firms to power special systems. e.g.: DYSEAC, SEAC, Lincoln TX-2, AMOEBA. [20][21]

## VI. CONCLUSION

In this paper we have presented the Heterogeneous Operating Systems with examples. We describe the information regarding operating system and issues or benefits of operating system so, it's a paper for awareness of operating system. Given the current state of the operating system market and the research field, GUI may be used to provide a bridge between both fields and promote the development of more flexible and cooperative operating systems. This would provide system administrators and programmers with the flexibility needed to develop user-friendly operating environments and applications that are not limited by the choice of a single OS.

# Appendix A

**Table 4 Comparison of different features of Operating system**

| System | Connectivity | Stability | Scalability | Multiuser | Multiplatform | POSIX | Non-Proprietary |
|---|---|---|---|---|---|---|---|
| Legacy System | Poor | Good | Medium-Huge | Yes | No | No | No |
| MS-DOS | None | Poor | Small | No | No | No | No |
| Windows 3.x | Poor | Poor | Small | No | No | No | No |
| Windows95 | SMB Only | Fair | Small | Insecure | No | No | No |
| WindowsNT | SMB+ | Fair | Small-Medium | Yes | Yes, 2 | Some | No |
| WindowXP | Excellent | Excellent | Small-Huge | Yes | Yes, Many | Yes | No |
| UNIX | Excellent | Excellent | Small-Huge | Yes | Yes, Many | Yes | No |
| Linux | Excellent | Excellent | Small-Huge | Yes | Yes, Many | Yes | No |

**Table 6 Comparison of technical Parameters of Operating System**

| Name | Architecture | File System support | Kernel Type | Source Line of code | GUI |
|---|---|---|---|---|---|
| FreeBSD | x86, x86-64, PC98, SPARC, others | UFS2, ext2, ext3, FAT, ISO 9660, UDF, NFS, ReiserFS (read only), XFS (experimental), ZFS and others | Monolithic with modules | | No |
| HP-UX | PA-RISC, IA-64 | VxFS, HFS, ISO 9660, UDF, NFS, SMBFS | Monolithic with modules | | No |
| IBM i | IBM | 1988 | OS/400 | | No |
| IRIX | SGI | 1988 | Unix system V | | No |
| Inferno | x86, PPC, SPARC, Alpha, MIPS, others | Styx/9P2000, kfs, FAT, ISO 9660 | Monolithic with modules, user space file systems | | Yes |
| Linux | x86, x86-64, PPC, SPARC, Alpha, others | ext2, ext3, ext4, ReiserFS, FAT, ISO 9660, UDF, NFS, and others | Monolithic with modules | 9 million lines of code | Yes |
| Mac OS Classic | 68k, PPC | HFS+, HFS, MFS (Mac OS 8.0 and before), AFP, ISO 9660, FAT(Sys 7 and later), UDF | Monolithic with modules | | Yes |
| Mac OS X | PPC, x86, x86-64, ARM | HFS+ (default), HFS, UFS, AFP, ISO 9660, FAT, UDF, NFS, SMBFS, NTFS (read only), FTP, WebDAV, ZFS (experimental | Hybrid | 86 millions | Yes |
| Mac OS X Server | Apple Inc. | HFS+ (default), HFS, UFS, AFP, ISO 9660, FAT, UDF, NFS, SMBFS, NTFS, FTP | Nextstep/ OPENSTEP/ MAC OS, UNIX | | Yes |
| OS/2 | x86 | HPFS, JFS, FAT, ISO 9660, UDF, NFS | Monolithic with modules | | No |
| DOS | x86 | FAT, | Monolithic | 45 million | Yes |
| Windows Server (NT Family) | x86, x86-64, IA-64 | NTFS, FAT, ISO 9660, UDF; 3rd-party drivers support ext2, ext3, reiserfs, and HFS | Hybrid | 45 million | Yes |
| Microsoft Window (Classic Family) | Microsoft | 1985 | Ms-Dos, Windows1 and later | | Yes |
| Microsoft Window(NT Family) | x86, x86-64 | NTFS, FAT exFAT ISO 9660, UDF; 3rd-party drivers support ext2, ext3, reiserfs, HFS+, FATX, and HFS | Hybrid | 40 million | Yes |
| Windows 2000 | IA-64, x86 | NTFS, FAT | Hybrid | | Yes |
| Windows XP | IA-32, x86-64 and Itanium | NTFS, FAT | Hybrid | | Yes |

**Table 5 Comparison Basic of Operating System**

| Name | Creator | First Public release | Predecessor | Latest stable version | Latest release date | Cost/Availability | Target System Type |
|---|---|---|---|---|---|---|---|
| **FreeBSD** | The FreeBSD Project | 1993 | 386BSD | 8.1 | 2010 | Free | Server, Workstation, NetApp, Embedded sys. |
| **HP-UX** | Hewlett-Packard | 1983 | Unix System V | 11.31"11iv3" | 2007 | Rs. 18272.56 | Server, Workstation |
| **IBM i** | IBM | 1988 | OS/400 | V6R1.1 | 2009 | Bundled with Hardware | Server |
| **IRIX** | SGI | 1988 | Unix system V | 6.5.30 | 2006 | Bundled with Hardware | Server, Workstation |
| **Inferno** | Bell Labs | 1997 | Plan 9 | Fourth Edition | 2007 | Free | Netapp, Server, Embedded System |
| **Linux** | Richard Stallman LinusTorvalds, Et.al | 1992 | Unix, Minux | Linux Kernel, GNU C library 2.11 | 2010 | Free | Just like linux |
| **Mac OS** | Apple Inc. | 1984 | None | 9.2.2 | 2002 | Bundled with 68K and Power PC macs | Workstation, Personal computer |
| **Mac OS X** | Apple Inc. | 2001 | Nextstep/ OPENSTEP/ MAC OS, UNIX | 10.6.6 | 2011 | Bundled with Hardware | Workstation, Personal comp., Embedded Sys. |
| **Mac OS X Server** | Apple Inc. | 2001 | Nextstep/ OPENSTEP/ MAC OS, UNIX | 10.6.4 | 2010 | Bundled with Hardware | Server |
| **DOS** | Microsoft | 1981 | 86 DOS/QDOS | 8.0 | 2000 | Bundled with Hardware | Workstation |
| **OS/2** | IBM & Microsoft | 1987 | Unix, Windows 3.x | 4.52 | 2001 | Rs. 13704.42 | Personal Computer, Server |
| **Windows Server (NT Family)** | Microsoft | 1993 | Ms-Dos, Os/2, Windows 3.x | Windows server R2(NT 6.1.7600) | 2009 | Rs 21424.58 | Server, Netapp, Embedded system, HPC |
| **Microsoft window (Classic Family)** | Microsoft | 1985 | Ms-Dos, Windows1 and later | Windows ME | 2000 | Outdated Product no longer for sale | Personal computer, Embedded system, Media center ,Tablet PC |
| **Microsoft Window(NT Family)** | Microsoft | 1983 | MS-DOS, OS/2, Windows 3.x | Windows 7(NT 6.1.7600) | 2009 | Rs.4565.85/ Home Basic | Workstation, Personal Computer, Media Center Tablet PC, Embedded |
| **Microsoft Windows 2000** | Microsoft | 2000 | Windows NT 4.0 | 5.0 (Build 2195: Service Pack 4) | 2005 | Rs. 14572.36 | Workstation, Personal Computer, Embedded |
| **Microsoft Window XP** | Microsoft | 2001 | Windows 2000, Windows Me | 5.1 (Build 2600: Service Pack 3) | 2008 | Rs.6500 /Home Basic | Workstation, Personal Computer, Media Center Tablet PC, Emb. |